\documentclass[onecolumn,showpacs,showkeys,preprintnumbers,amsmath,amssymb]{revtex4}
\usepackage{graphicx}
\usepackage{dcolumn}
\usepackage{bm}
\usepackage{epsfig}
\usepackage{rotating}

\begin{document}
\title{Hexagonal type Ising nanowire with spin-1 core and spin-2 shell structure}

\author{Mehmet Erta\c{s} and Ersin Kantar$*$
\\
Department of Physics, Erciyes University, 38039 Kayseri, Turkey
}

\altaffiliation[]{Corresponding Author: Department of Physics, Erciyes University, 38039 Kayseri, Turkey
\\Email: ersinkantar@erciyes.edu.tr (E. Kantar);
\\Tel.: +90 352 4374938x33136;
\\Fax: +90352 4374931.}

\begin{abstract}Thermodynamic properties and phase diagrams of a mixed spin-(1, 2) Ising ferrimagnetic system with single ion anisotropy on hexagonal nanowire are studied by using effective-field theory with correlations. The susceptibility, internal energy and specific heat of the system are numerically examined and some interesting phenomena in these quantities are found. The effect of the Hamiltonian parameters on phase diagrams are examined in detail. Besides second-order phase transition, lines of first-order transition and tricritical points are found. In particular, we found that for some negative values of single-ion anisotropies, there exist first-order phase transitions.
\end{abstract}

\maketitle

\section{Introduction}
In recent years, nanoparticles, nanodots, nanofilms, nanorods, nanobelts, nanowires and nanotubes have attracted a great interest in scientific research \cite{1,2,3,4,5,6}. These magnetic nanomaterial have many typical, peculiar and unexpected physical properties compared with those in bulk materials \cite{7}. Moreover, they can be used in biomedical applications \cite{8,9}, sensors \cite{10}, nonlinear optics \cite{11}, permanent magnets \cite{12}, environmental remediation \cite{13}, and information storage devices \cite{14,15,16}.

On the other hand, theoretically, magnetic properties of nanomaterial have been investigated successfully by adopting a core-shell structure of the Ising systems. In this context, by means of various techniques such as mean-field theory (MFT) \cite{17,18}, effective-field theory (EFT) with correlations \cite{19,20,21,22,23,24,25,26,27,27a,27b,27c,27d,27e}, and Monte Carlo (MC) simulations \cite{28,29,30,31} have been studied magnetic behaviors of nanomaterial. Furthermore, the dynamic magnetic properties of nanomaterials with core/shell structure have been investigated by means of effective-field theory based on the Glauber-type stochastic (DEFT) \cite{d28,d29,d30,d31}.

We should also mention that mixed spin Ising systems have attracted a great deal of attention. The systems provide simple models which can show ferrimagnetic behavior and the systems have less translational symmetry than their single spin counterparts; hence exhibit many new phenomena that cannot be observed in the single-spin Ising systems. These systems have been studied extensively by using the well-known methods in equilibrium statistical physics. The well-known mixed spin Ising systems are (1/2, 1) \cite{32,33,34,35,27d}, (1/2, 3/2) \cite{37,38,27e}, (1, 3/2) \cite{40,41,42,43} and (2, 5/2) \cite{44,45,46} Ising systems. Furthermore, the nonequilibrium aspects of mixed spins (1, 1/2) \cite{47,48}, (1/2, 3/2) \cite{49,50}, (1, 3/2) \cite{51}, (2, 5/2) \cite{52}, have been also investigated. Up until now, few people have ever touched upon a mixed Ising model with both spin-integer ions, namely the mixed spin (1, 2) Ising system. \cite{53,54,55,56,57,58,59,60,61}. Despite these studies, magnetic properties of a mixed spin (1, 2) on hexagonal Ising nanowire (HIN) have not been studied. To this end, in this paper, the effects of the interfacial coupling and the interaction coupling in the surface as well as crystal field, on the thermodynamic properties and phase diagrams of HIN system with core/shell structure are discussed within the framework of the EFT with correlations.

The paper is organized as follows. In Section 2, formulation of the model and its effective-field solution is presented briefly. The detailed numerical results and discussions are given in Section 3. Finally, Section 4 is devoted to a summary and a brief conclusion.

\section{Formulation of the Model and Its Effective-Field Solution}

The Hamiltonian of the HIN system includes nearest-neighbor interactions and the crystal field is given as follows

\begin{equation}
H=-J_{S} \sum \limits _{\left\langle ij\right\rangle }S_{i}^{}  S_{j}^{} -J_{C} \sum \limits _{\left\langle mn\right\rangle }\sigma _{m}^{}  \sigma _{n}^{} -J_{1} \sum \limits _{\left\langle im\right\rangle }S_{i}^{}  \sigma _{m}^{} -\Delta \left(\sum \limits _{i}\left(S_{i}^{} \right)^{2}  +\sum \limits _{m}\left(\sigma _{m}^{} \right)^{2}  \right)-h\left(\sum \limits _{i}S_{i}^{}  +\sum \limits _{m}\sigma _{m}^{}  \right),
\end{equation}

where $\sigma =\pm 1,\, 0$and S= $\pm$2, $\pm$1, 0. $<$\dots $>$ indicates summation over all pairs of nearest-neighbor sites. The exchange interaction parameters J${}_{S}$, J${}_{C}$ and J${}_{1}$ are interactions between the two nearest-neighbor magnetic atoms at the shell, core and between the shell and core, respectively (see Fig. 1). $\Delta$ stands for the single-ion anisotropy, i.e. the crystal field, and h is the external magnetic field. The surface exchange interaction $J_{S} =J_{C} \left(1+\Delta _{S} \right)$ and interfacial coupling $r=J_{1} /J_{C} $ are often defined to clarify the effects of the surface and interfacial exchange interactions on the physical properties in the nanosystem, respectively.

Within the EFT with correlations framework, one can easily find the magnetizations ($m_{S} $ and $m_{C} $), the quadruple moments ($q_{S} $ and $q_{C} $) , the octupolar moment ($r_{S} $) and hexadecapole moment ($v_{S} $) as coupled equations, for the HIN system as follows:

\begin{subequations}
\begin{equation}
\left\{\begin{array}{l} {m_{S} } \\ {q_{S} } \\ {r_{S} } \\ {v_{S} } \end{array}\right\}=\left[{\rm a}_{{\rm 0}} {\rm +a}_{{\rm 1}} m_{S} {\rm +a}_{{\rm 2}} q_{S} {\rm +a}_{{\rm 3}} r_{S} {\rm +a}_{{\rm 4}} v_{S} \right]^{4} \left[1+m_{C} sinh \left({\rm J}_{{\rm 1}} \nabla \right)+q_{C}^{} \left(cosh \left({\rm J}_{{\rm 1}} \nabla \right)-1\right)\right]\left[\left\{\begin{array}{l} {{\rm F}_{{\rm m}} \left. \left({\rm x}\right)\right|_{{\rm x=0}} } \\ {{\rm F}_{{\rm q}} \left. \left({\rm x}\right)\right|_{{\rm x=0}} } \\ {{\rm F}_{{\rm r}} \left. \left({\rm x}\right)\right|_{{\rm x=0}} } \\ {{\rm F}_{{\rm v}} \left. \left({\rm x}\right)\right|_{{\rm x=0}} } \end{array}\right\}\right],
\end{equation}
\begin{equation}
\left\{\begin{array}{l} {m_{C} } \\ {q_{C} } \end{array}\right\}=\left[1+m_{C} sinh \left({\rm J}_{{\rm C}} \nabla \right)+q_{C}^{} \left(cosh \left({\rm J}_{{\rm C}} \nabla \right)-1\right)\right]^{2} \left[{\rm b}_{{\rm 0}} {\rm +b}_{{\rm 1}} m_{S} {\rm +b}_{{\rm 2}} q_{S} {\rm +b}_{{\rm 3}} r_{S} {\rm +b}_{{\rm 4}} v_{S} \right]^{6} \left[\left\{\begin{array}{l} {{\rm G}_{{\rm m}} \left. \left({\rm x}\right)\right|_{{\rm x=0}} } \\ {{\rm G}_{{\rm q}} \left. \left({\rm x}\right)\right|_{{\rm x=0}} } \end{array}\right\}\right]\,
\end{equation}
\end{subequations}

where the $a_{i} \, and\, b_{i}$ coefficients are given in the Appendix. The functions $F\left(x\right)$ and ${\rm G}\left(x\right)$ are defined as

\begin{subequations}
\begin{equation}
F_{m} (x)=\frac{1}{2} \frac{4\, sinh \left[2\beta \left(x+h\right)\right]+2\, sinh \left[\beta \left(x+h\right)\right]\, \, exp \left(-3\beta \, \Delta \right)}{\, cosh \left[2\beta \left(x+h\right)\right]+cosh \left[\beta \left(x+h\right)\right]\, \, exp \left(-3\beta \, \Delta \right)+\, \, exp \left(-4\beta \, \Delta \right)},
\end{equation}
\begin{equation}
F_{q} (x)=\frac{1}{2} \frac{8\, cosh\left[2\beta\left(x+h\right)\right]+2\, cosh\left[\beta\left(x+h\right)\right]\, \, exp\left(-3\beta\, \Delta\right)}{\, cosh\left[2\beta\left(x+h\right)\right]+cosh\left[\beta\left(x+h\right)\right]\, \, exp\left(-3\beta\, \Delta\right)+\, \, exp\left(-4\beta\, \Delta\right)}
\end{equation}
\begin{equation}
{ F}_{{ r}} { (x)=}\frac{{ 1}}{{ 2}} \frac{{ 16}\, { sinh}\left[{ 2 \beta}\left({ x+h}\right)\right]{ +2}\, { sinh}\left[{ \beta}\left({ x+h}\right)\right]\, \, { exp}\left({ -3 \beta}\, { \Delta}\right)}{\, { cosh}\left[{ 2 \beta}\left({ x+h}\right)\right]{ +cosh}\left[{ \beta}\left({ x+h}\right)\right]\, \, { exp}\left({ -3 \beta}\, { \Delta}\right){ +}\, \, { exp}\left({ -4 \beta}\, { D}\right)}
\end{equation}
\begin{equation}
F_{v} (x)=\frac{1}{2} \frac{32\, cosh \left[2\beta \left(x+h\right)\right]+2\, cosh \left[\beta \left(x+h\right)\right]\, \, exp \left(-3\beta \, \Delta \right)}{\, cosh \left[2\beta \left(x+h\right)\right]+cosh \left[\beta \left(x+h\right)\right]\, \, exp \left(-3\beta \, \Delta \right)+\, \, exp \left(-4\beta \, \Delta \right)}
\end{equation}
\begin{equation}
G_{m} (x)=\frac{2\; sinh[\beta (x+h)]}{2\; cosh[\beta \; (x+h)]+exp(-\beta \; \Delta)}
\end{equation}
\begin{equation}
G_{q} (x)=\frac{2{ \; }cosh [\beta (x+h)]}{2{ \; }cosh [\beta { \; (}x+h)]+exp (-\beta { \; }\Delta )}
\end{equation}
\end{subequations}

Here, $\beta ={1\mathord{\left/ {\vphantom {1 k_{{ B}} T}} \right. \kern-\nulldelimiterspace} k_{{ B}} T} $, $T$ is the absolute temperature, k${}_{B}$ is the Boltzmann constant and k${}_{B}$ = 1.0 throughout the paper. By using the definitions of the magnetizations in Eqs. (2a) and (2b), the total magnetization $M_{T} $ of each site can be defined from Fig. 1 as $M_{T} =1/7\left(\, 6m_{S} +m_{C} \right).$

We should also mention that we did not examine the thermal behaviors of$q_{S} $, $q_{C} $, $r_{S} $ , and $v_{S} $due to the reason that our Hamiltonian did not contain the biquadratic exchange interaction parameter as seen in Eq. (1). However, we need Eqs. (2a) and (2b) to determine the behaviors of $M_{S} $ and$M_{C} $.

The internal energy of per site of the system can be calculated as

\begin{equation}
\frac{U}{N} =-\frac{1}{2} \left(\langle U_{C} \rangle \, +\, \langle U_{S} \rangle \right)-\, D\left(\langle q_{C} \rangle \, +\, \langle q_{S} \rangle \right)\, -\, h\left(\langle m_{C} \rangle \, +\, \langle m_{S} \rangle \right),
\end{equation}

where,

\begin{subequations}
\begin{equation}
U_{S} =\frac{\partial }{\partial \nabla } \left[{ a}_{{ 0}} { +a}_{{ 1}} m_{S} { +a}_{{ 2}} q_{S} { +a}_{{ 3}} r_{S} { +a}_{{ 4}} v_{S} \right]^{4} \left[1+m_{C} sinh \left({ J}_{{ 1}} \nabla \right)+q_{C}^{} \left(cosh \left({ J}_{{ 1}} \nabla \right)-1\right)\right]{ F}_{{ 1}} \left. \left({ x+h}\right)\right|_{{ x=0}}
\end{equation}
\begin{equation}
U_{C} =\frac{\partial }{\partial \nabla } \left[1+m_{C} sinh \left({ J}_{{ C}} \nabla \right)+q_{C}^{} \left(cosh \left({ J}_{{ C}} \nabla \right)-1\right)\right]^{2} \left[{ b}_{{ 0}} { +b}_{{ 1}} m_{S} { +b}_{{ 2}} q_{S} { +b}_{{ 3}} r_{S} { +b}_{{ 4}} v_{S} \right]^{6} { G}_{{ 1}} \left. \left({ x+h}\right)\right|_{{ x=0}}
\end{equation}
\end{subequations}

The specific heat of the system can be obtained from the relation

\begin{equation} \label{GrindEQ__6_}
C=\frac{\partial }{\partial T } \left(\frac{\partial U}{\partial T} \right).
\end{equation}

On the other hand, we will investigate the thermal behavior of the susceptibility \textit{$\chi $${}_{\alpha }$} for the system with a crystal-field. The susceptibility for the system can be determined easily from the following equation:

where \textit{$\alpha $} (\textit{$\alpha $} = \textit{$m_{S}^{} \, and\, \, m_{C}^{} $}) is taken the values of the longitudinal magnetizations. By using Equations (2a), (2b) and (6), we can easily obtain the susceptibilities \textit{$\chi _{S}^{} \, \, \, and\, \, \chi _{C}^{} $ }as follows:

\begin{subequations}
\begin{equation}
\chi _{S} =A_{1} \, \chi _{S} +A_{2} \, \chi _{C} +A_{3} \, \frac{\partial F_{1} \left(x\right)}{\partial h}
\end{equation}
\begin{equation}
\chi _{C} =B_{1} \, \chi _{C} +B_{2} \, \chi _{S} +B_{3} \, \frac{\partial G_{1} \left(x\right)}{\partial h}
\end{equation}
\end{subequations}

Here, A${}_{i }$ and B${}_{i}$ (i=1, 2 and 3) coefficients have complicated and long expressions, hence they will not give. The total susceptibilities of per site can be obtain via $\chi _{T} ={1\mathord{\left/ {\vphantom {1 7}} \right. \kern-\nulldelimiterspace} 7} \, \left(6\, \chi _{S} +\, \chi _{C} \right)$.

\section{Numerical results and discussions}
In this section, we investigate some interesting and typical results of the HIN system with core/shell structure. Some thermodynamic and magnetic quantities (magnetization, susceptibility, internal energy and specific heat) of the HIN system with core/shell structure are studied and discussed for selected values of the interaction parameters. Moreover, we also present the phase diagrams of the system, seen in Fig. 4 and 5.

\subsection{Thermodynamic properties}
 We examine the thermodynamic properties, such as magnetization, susceptibility, internal energy and specific heat of the system. Firstly, we have fixed \textit{J${}_{C }$}= 1.0 throughout of the paper.

\subsubsection{Magnetization curves and susceptibilities}
Thermal behaviors of the magnetizations and the corresponding susceptibilities $\chi$ (core, shell and total) are illustrated in Fig. 2. We obtained the thermal behaviors of magnetizations and the corresponding susceptibilities by solving Eqs. (2a) and (2b), and Eqs. (2a) and (2b) and (7), respectively. We present a few representative graphs to display their behaviors for first and second-order phase transitions the HIN, as shown in Figs. 2(a) - (d). The results are depicted in Figs. 2(a) - (b) for the values of ferrimagnetic interface coupling r $>$ 0, $\Delta{}_{S}$=0, $\Delta$=1, and antiferromagnetic interface coupling r $<$ 0, $\Delta{}_{S}$=0, $\Delta$=1, respectively. In Figs. 2(a) and (b), T${}_{C}$ represent the second-order phase transition temperatures from the ferrimagnetic (1, 2) and antiferromagnetic (-1, 2) phase to the P phase, respectively. From these figures are seen that the core/shell and total magnetizations go to zero continuously as the temperature increases and a second-order phase transition occurs at same critical temperature value, T${}_{C}$= 10.12. When the temperature approaches T${}_{C}$, the susceptibilities c increase very rapidly and go to infinity at T${}_{C}$ = 10.12. Figs. 2(c) and (d) illustrate the thermal variation of the magnetizations and susceptibilities for the values of ferrimagnetic interface coupling r $>$ 0, $\Delta{}_{S}$=0, $\Delta$=-3, and antiferromagnetic interface coupling r $<$ 0, $\Delta{}_{S}$=0, $\Delta$=-3, respectively. In Figs. 2(c) and (d), T${}_{t}$=0.84 represents the first-order phase transition temperatures from the ferrimagnetic (1, 2) and antiferromagnetic (-1, 2) phase to the P phase, respectively. Figures 2(c) and 2(d) show core/shell magnetizations go to zero discontinuously as the temperature increases; hence, a first-order phase transition occurs at same critical temperature value, T${}_{t}$= 0.84. Moreover, in the vicinity of T, the core/shell susceptibility c rapidly increases for T $<$ T${}_{t}$ and suddenly decreases for T $>$ T${}_{t}$.

\subsubsection{Internal Energies and Specific Heats}
We will also investigate the temperature variations of the internal energy and specific heat to study the present system in detail. Let us study the thermal variation of the internal energy and specific heat of the system for the HIN by solving Eqs. (2a), (2b), (4) and (6) numerically. In the case of r $>$ 0, we have plotted the temperature dependence of the internal energy and specific heat for the selected fixed values of $\Delta{}_{S}$=0, $\Delta$=1 in Fig. 3(a). Figure 3(a) shows that when T $<$ T${}_{C}$, the specific heat for the second-order phase transition increases with increasing temperature. The specific heats may express the discontinuity at T = T${}_{C}$, although in the high-temperature region (T $>$ T${}_{C}$ ), they take finite values. Moreover, the maxima of the specific heat correspond to points where the first-order derivatives of the internal energy U are discontinuous, at which the second-order phase transitions occur. In Fig. 3(b), for the case of r $<$ 0, we have presented the temperature dependence of the internal energy and specific heat for the selected fixed values of $\Delta{}_{S}$=0, $\Delta$=-3. Figure 3(b) shows that the specific heat arrives at a maximum value and reduces suddenly to a small value at the first-order phase transition temperature. The corresponding internal energies are given in Fig. 3(d). In that figure, the internal energies increase with the increasing temperature discontinuously at the first-order phase transition temperature.

\subsection{Phase Diagrams}
In this subsection, we shall show some typical results for the mixed spin (1, 2) Ising model with a crystal field. We have obtained the phase diagrams two different planes, namely ($\Delta $, T) and (r, T) for hexagonal nanowire.

\subsubsection{Phase Diagrams in (${\mathbf \Delta }$, T) plane}
At first, we present the phase diagrams of the model in the ($\Delta $, T) plane, illustrated in Fig. 4. In these phase diagrams, the solid and dashed lines represent the second- and first-order phase transition lines, respectively, and the tricritical points are denoted by filled circles. It is clear that the second- and first-order phase transition lines separate the antiferromagnetic and ferrimagnetic phases from the paramagnetic (p) phase. From these phase diagram the following phenomena have been observed. (i) Each one of the phase diagrams exhibits only one tricritical point where the second-order phase transition turns to a first-order one. (ii) The grey triangles correspond to the multicritical points, and these triangles separate the ferromagnetic (1, 1) phase from the ferrimagnetic (1, 2) phase. The similarly behavior is observed the Fig. 2 of the Ref. \cite{59}. (iii) The reentrant behavior exists in the HIN system, i.e., the system will be disordered (paramagnetic) phase at very low temperatures and as the temperature increases the system ordered phase at a critical temperature T${}_{C1}$ and finally the paramagnetic phase at a higher critical temperature T${}_{C2}$. The phenomenon may be caused by the competition between the exchange interaction and negative single-ion anisotropy strength.

\subsubsection{Phase Diagrams in (r, T) plan}
In Fig. 5, the phase diagram of mixed spin-(1, 2) HIN system is obtained to examine the influence of the interfacial coupling. Fig. 5 shows the variations of T as a function of r, when the parameter $\Delta$ is fixed at $\Delta$=0 and the values of $\Delta{}_{S}$ are changed ($\Delta{}_{S}$ =$-$0.9, -0.5 and 0.0). In Fig. 5, the phase transition region is divided into two phases namely p and i. In this figure, we can see that i phase become larger with the increasing absolute (r) and $\Delta{}_{S}$. From Fig. 5, we can see that all the phase transition is the second-order phase transition.

\section{Summary and Conclusions}
In this paper, we have studied the magnetizations, susceptibilities and internal energies of mixed spin-(1, 2) HIN system on by using the effective-field theory with correlations. We also investigate the phase diagrams, seen in Fig. 4 and 5. It has been shown that the system both a second-order phase transition and first-order phase transition. Moreover, from the phase diagrams, reentrant phenomena can be found under certain conditions of the Hamiltonian parameters. Finally, we hope that the study of HIN system may open a new ferrimagnetism as well as new field in the research of magnetism and present work will be potentially helpful for studying higher spins and more complicated nanowire systems.

\section{Appendix}

\[\left|\begin{array}{l} {a_{0} =1\, \, ,} \\ {a_{1} =\frac{1}{6} \, \left[8\, sinh \left({ J}_{{ S}} \nabla \right)-\, sinh \left(2\, { J}_{{ S}} \nabla \right)\right]\, \, ,} \\ {a_{2} =\frac{1}{12} \, \left[16cosh \left({ J}_{{ S}} \nabla \right)-\, cosh \left(2\, { J}_{{ S}} \nabla \right)-15\right],\, \, \, \, \, } \\ {a_{3} =\frac{1}{6} \, \left[\, sinh \left(2\, { J}_{{ S}} \nabla \right)-2\, sinh \left({ J}_{{ S}} \nabla \right)\right]\, \, ,} \\ {a_{4} =\frac{1}{12} \, \left[cosh \left(2\, { J}_{{ S}} \nabla \right)-4\, cosh \left({ J}_{{ S}} \nabla \right)+3\right]\, \, ,} \end{array}\right|\, \, \, and\, \, \, \left|\begin{array}{l} {b_{0} =1\, \, ,} \\ {b_{1} =\frac{1}{6} \, \left[8\, sinh \left({ J}_{{ 1}} \nabla \right)-\, sinh \left(2\, { J}_{{ 1}} \nabla \right)\right]\, \, ,} \\ {b_{2} =\frac{1}{12} \, \left[16cosh \left({ J}_{{ 1}} \nabla \right)-\, cosh \left(2\, { J}_{{ 1}} \nabla \right)-15\right],\, \, \, \, \, } \\ {b_{3} =\frac{1}{6} \, \left[\, sinh \left(2\, { J}_{{ 1}} \nabla \right)-2\, sinh { J}_{{ 1}} \nabla \right]\, \, ,} \\ {b_{4} =\frac{1}{12} \, \left[cosh \left(2\, { J}_{{ 1}} \nabla \right)-4\, cosh \left({ J}_{{ 1}} \nabla \right)+3\right]\, \, .} \end{array}\right|\]

\begin{center}
\textbf{List of the Figure Captions}
\end{center}

\begin{itemize}
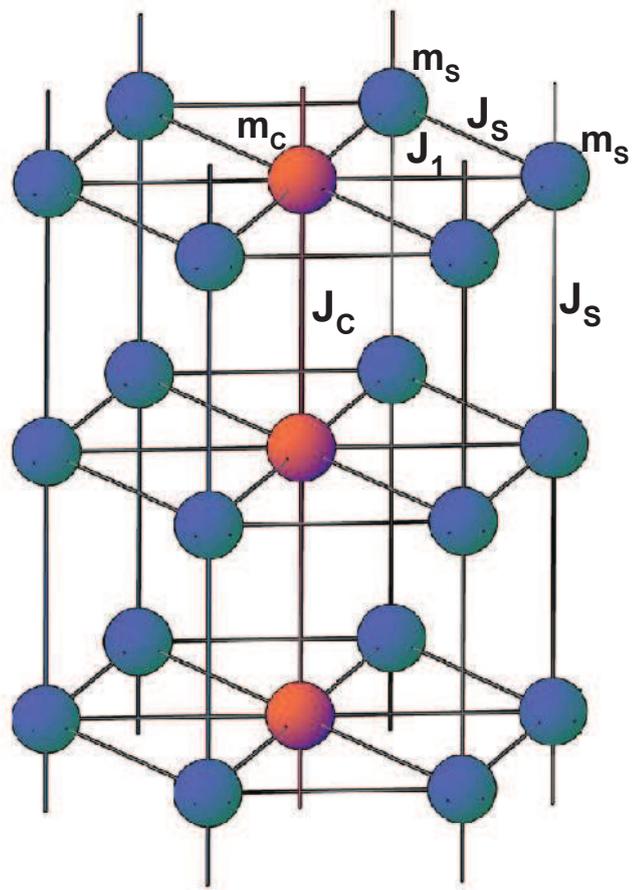
\item [\textbf{Fig. 1}] (Color online) Schematic representation of hexagonal Ising nanowire. The blue and red spheres indicate magnetic atoms at the surface shell and core, respectively.

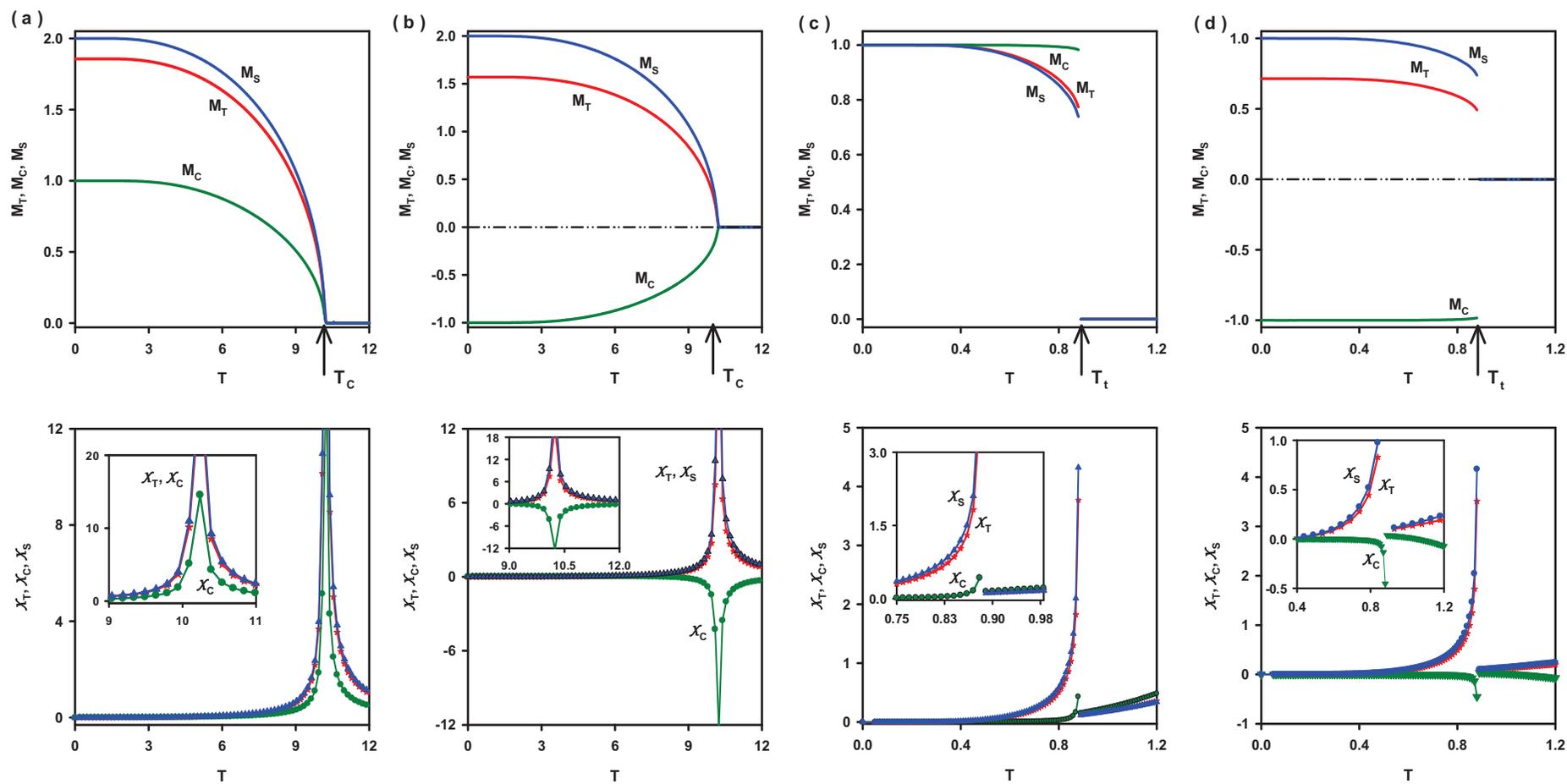
\item [\textbf{Fig. 2}] (Color online) The behavior of the core, shel and total magnetizations and magnetic susceptibilities as a function of temperature on the hexagonal nanowire. TC (thick arrow) is the second-order phase transition temperature from the ferrimagnetic phase to the paramagnetic phase. Tt (dashed arrow) represents the first-order phase transition temperature from the antiferromagnetic phase to the paramagnetic phase. (\textbf{a}) $J_{1}=1,  J_{C}=1, \Delta_{S}=0, \Delta=1$ (\textbf{b}) $J_{1}=-1, J_{C}=1, \Delta_{S}=0, \Delta=1$ (\textbf{c})$J_{1}=1,  J_{C}=1, \Delta_{S}=0, \Delta=-3$ (\textbf{d}) $J_{1}=-1, J_{C}=1, \Delta_{S}=0, \Delta=-3$

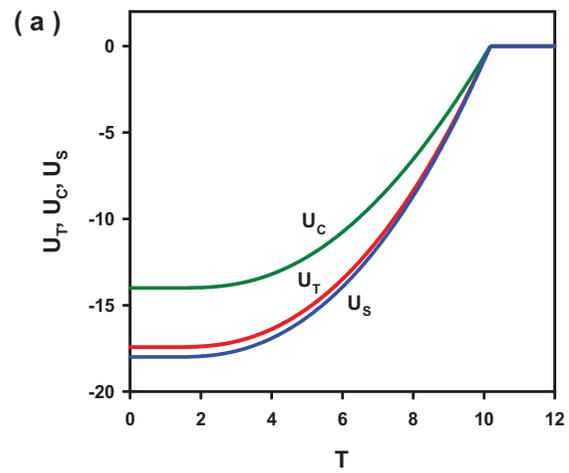
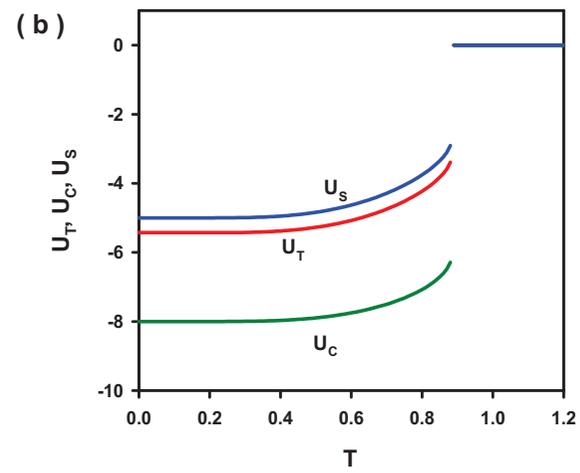
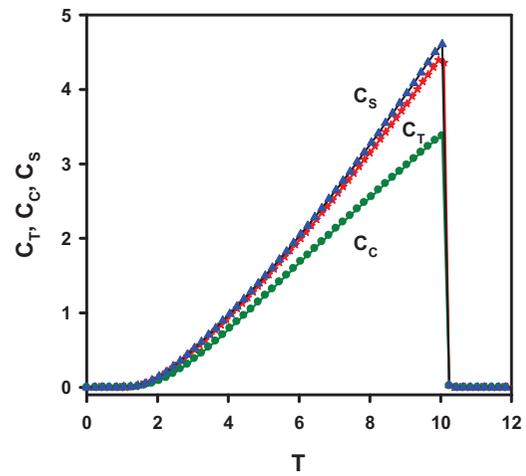
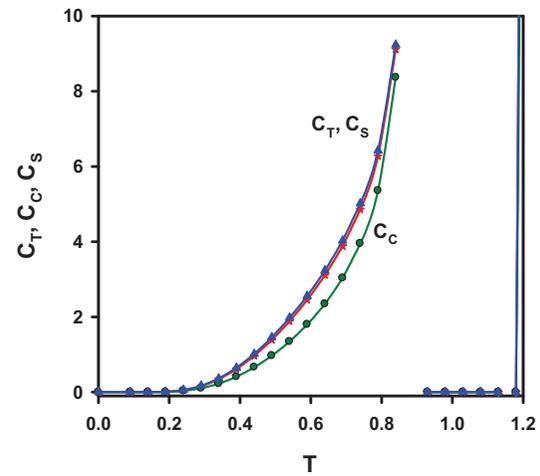
\item [\textbf{Fig. 3}] (Color online) Temperature dependence of the specific heats and internal energies for the model on the hexagonal nanowire.
(\textbf{a}) $J_{1}=1, J_{C}=1, \Delta_{S}=0, \Delta=1$(\textbf{b}) $J_{1}=1, J_{C}=1, \Delta_{S}=0, \Delta=-3$

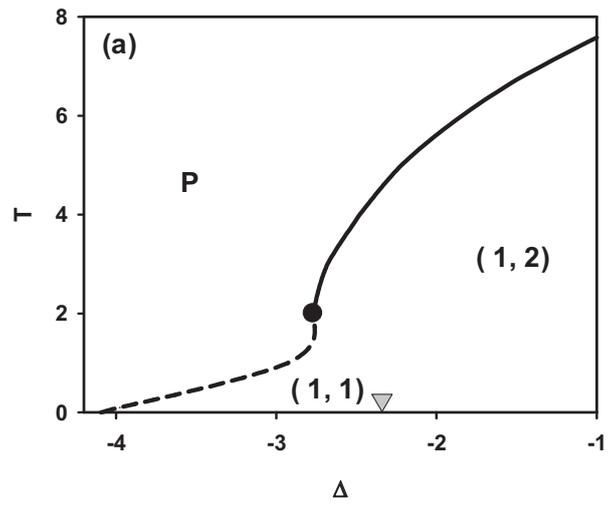
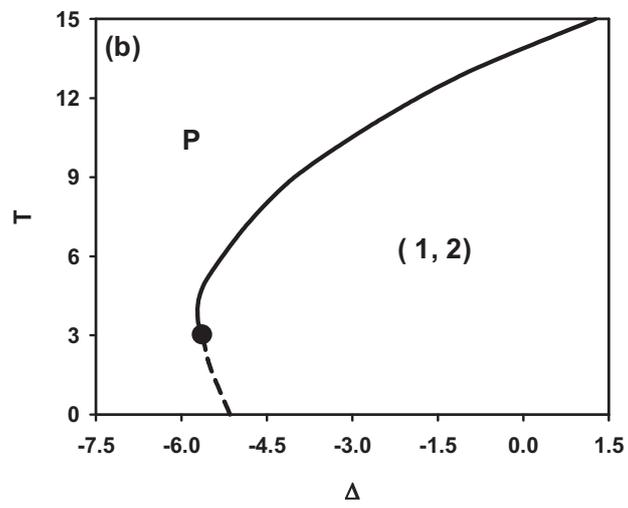
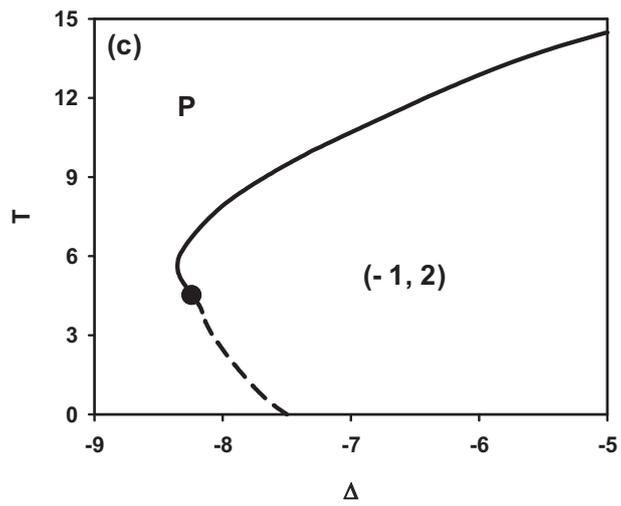
\item [\textbf{Fig. 4}] Phase diagrams in the ($\Delta, T$) plane for the mixed spin Ising model consisting of spins σ=1 and S=2 within the EFT at zero longitudinal magnetic field. The solid and dashed lines represent the second- and first-order phase transition lines, respectively; the tricritical points are denoted by filled circles; the  grey triangles correspond to the multicritical points and the triangle separate the ordered phase from the disordered phase. (\textbf{a})$J_{1}=1,  J_{C}=1, \Delta_{S}=0$

    (\textbf{b})$J_{1}=3,  J_{C}=1, \Delta_{S}=0$(\textbf{c})$J_{1}=-3,  J_{C}=1, \Delta_{S}=0$

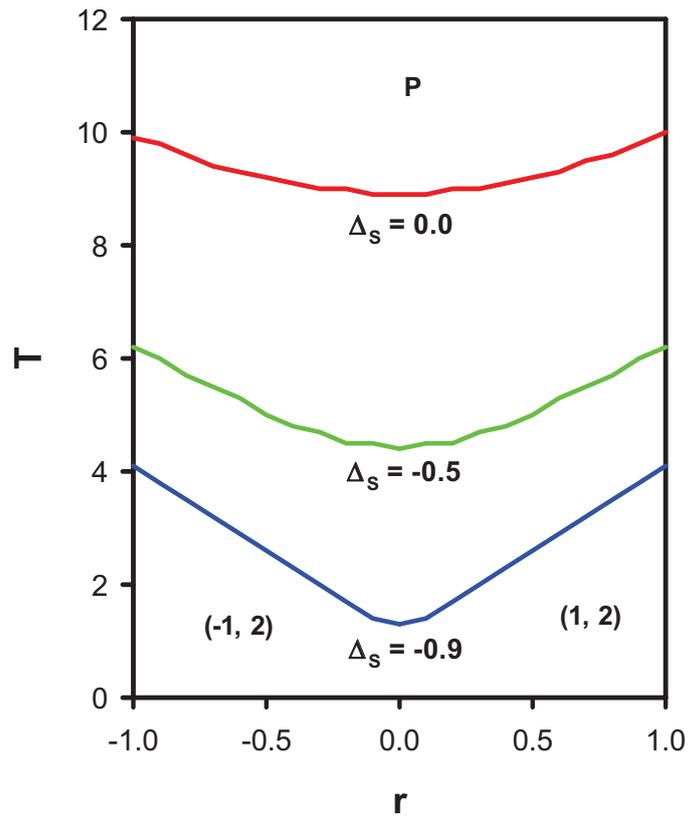
\item [\textbf{Fig. 5}] (Color online) Phase diagrams in the ($r, T$) plane for the mixed spin-1 and spin-2 hexagonal Ising nanowire. The solid line represent the second-order phase transition line. $\Delta$ is fixed at $\Delta$=0 and the values of $\Delta{}_{S}$ are changed ($\Delta{}_{S}$ =$-$0.9, -0.5 and 0.0).

\end{itemize}

\end{document}